# High Frequency top-down Junction-less Silicon Nanowire Resonators

*Alexandra Koumela[†], Sébastien Hentz[†], Denis Mercier[†], Cécilia Dupré[†], Eric Ollier[†],*

*Philip X.-L. Feng[§], Stephen T. Purcell[‡] and Laurent Duraffourg[†,*]*

[†]CEA, LETI, MINATEC Campus, 17 rue des Martyrs, 38054 GRENOBLE Cedex 9, France.

[§]Electrical Engineering, Case School of Engineering, Case Western Reserve University, Cleveland, OH 44106, USA.

[‡]Laboratoire de Physique de la Matière Condensée et Nanostructures, Université Lyon I, CNRS, UMR 5586, Domaine Scientifique de la Doua, F-69622 Villeurbanne Cedex, France.

[*]Corresponding Author, Email: laurent.duraffourg@cea.fr

ABSTRACT

We report here the first realization of top-down silicon nanowires (SiNW) transduced by both junction-less field effect transistor (FET) and the piezoresistive (PZR) effect. The suspended SiNWs are among the smallest top-down SiNWs reported to date, featuring widths down to ~20nm. This has been achieved thanks to a 200mm-wafer-scale, VLSI process fully amenable to monolithic CMOS co-integration. Thanks to the very small dimensions, the conductance of the silicon nanowire can be controlled by a nearby electrostatic gate. Both the junction-less FET and the previously demonstrated PZR transduction have been performed *with the same SiNW*. These self-transducing schemes have shown similar signal-to-background ratios, and the PZR transduction has exhibited a relatively higher output signal. Allan deviation ($\sigma_A$) of the same SiNW has been measured with both schemes, and we obtain $\sigma_A$~20ppm for the FET detection and $\sigma_A$~3ppm for the PZR detection at room temperature and low pressure. Orders of magnitude improvements are expected from tighter electrostatic control via changes in geometry and doping level, as well as from CMOS integration. The compact, simple topology of these elementary SiNW resonators opens up new paths towards ultra-dense arrays for gas and mass sensing, time keeping or logic switching systems in SiNW-CMOS platform.

# Introduction

During the last decade, nanoelectromechanical systems (NEMS) have shown great promise for both fundamental science and applications, providing new tools for studying quantum physics[1,2] including electromechanics and optomechanics[3]. Technological applications of NEMS are also emerging: these include integrated frequency clocks[4], logic switches[5], mixer filters[6], ultra-sensitive force detectors[7] and mass sensors[8]. Keys to real-world implementations though is a scalable, integrated, low-power transduction scheme, associated to a fabrication process allowing for individually addressable devices in ultra-large, ultra-dense arrays. Essential bricks of the puzzle have started to fall into place in recent years: VLSI-compatible transduction schemes have been proposed[9,10] for individual devices, and their implementation in real systems has been demonstrated for gas sensing[11] as well as for real-time single-molecule mass spectrometry[12]. Nevertheless, these NEMS include a fairly large number of electrical contacts per device, making their individual addressing complex when in arrays. Even though a major step forward was accomplished with the first realization of LSI NEMS arrays in collective addressing fashion[13], a transduction and readout scheme allowing simple, extremely compact and smart NEMS pixels is yet to be developed. Recently an elegant two-source, piezoresistive self-transducing scheme has been used with a carbon nanotube (CNT) and has shown sufficient signal-to-noise ratio (SNR) for sensitive detection[14]. A one-source field-effect transduction of CNTs has shown enough gain to demonstrate a few Dalton to a few tens of Dalton mass resolution[8]. The bottom-up process fabrication of the two latter cases has not allowed for scalability so far. On the other hand, scalable nanomechanical transductions have been proposed recently[15] [16] where a fin field effect transistor and a first type of junction-less field effect are embedded into the NEMS. However, the top-down processes used in these studies to obtain the necessary effects require localized doping with advanced lithography and additional mask levels.

In this paper, we introduce a low-power, scalable, top-down fabrication process with simple full sheet doping (i.e. one single dopant concentration for the suspended SiNW and the control gate), and minimum number of mask levels. We demonstrate the simple implementation of a junction-less field-effect transduction (only enabled by the minimum feature size achieved, below 30nm) as well as a piezoresistive (PZR) self-transducing scheme, on the same device. Both detection schemes are carefully compared: electromechanical resonances obtained thanks to the down-mixing technique[17] and frequency stabilities of the same SiNW have been measured with both schemes at room temperature and low pressure.

This comprehensive demonstration paves the way towards integrated low-cost, low-power, ultra-dense, ultra-large arrays of highly sensitive NEMS for various real-world applications.

## Device Operation and Fabrication Process

We report here a self-transducing NEMS device based on a suspended silicon nanowire resonator (SiNW), see Figure 1a). Each device comprises a doubly clamped beam electrically connected on both ends, as well as two lateral electrodes (for both actuation and detection). The mechanical motion is detected by monitoring the output current produced by conductance modulation of the SiNW. This modulation is obtained via two different principles: tension-induced piezoresistivity and junction-less field effect where the SiNW is itself a monolithic suspended transistor. Both principles are made possible thanks to intrinsic properties of silicon, ubiquitous in the standard microelectronics industry and its associated tools. These tools allow for NEMS with widths from 20 to 40nm and lengths on the order of a few microns, with measured resonances from 30MHz to 150MHz (see Figure 1b) to d)).

The fabrication process is based on CMOS-compatible techniques in order to facilitate future co-integration with electronic circuitry on the same substrate. The initial wafer is a silicon-on-insulator (SOI) wafer, which facilitates the release of the mechanical structure. Firstly the Si film of the SOI wafer is homogenously p-type implanted with boron ($N_a=10^{19}$ cm$^{-3}$) and then hybrid e-beam / DUV (deep ultra violet) lithography is used for the patterning of both nanowires and pads. Next etching of the devices and release of the structures with (hydrogen fluoride) are performed. In order to passivate the SiNW a thermal oxide of about 10nm is grown around the whole NW. Before the metallization of the contact pads, a layer of poly-Si is deposited to protect the release devices. The poly-Si is etched and AlSi is deposited to make the pads. The process flow is similar to the CMOS process presented elsewhere[18]. The final step is a second release of the resonators with etching of the poly-Si. The initial Si film is reduced down to 30nm by thermal oxidation, while the thermal oxide covers all around the structure. The nanowire crystal is oriented <110>. It should be noted that both nanowire and actuation electrodes are doped at the same initial level ($N_a=10^{19}$ cm$^{-3}$) with Boron. In both cases the effective dopant concentrations in the SiNWs have been found lower than the full sheet doping level. From resistance measurements the effective doping has been found to vary with the NW width, with lower values for thinner NWs. The doping concentration is $3.8 \times 10^{17}$cm$^{-3}$ for thin SiNWs ($w$=35nm), while for thicker nanowires ($w$=80nm) the extracted effective dopant concentration is $\sim 10^{18}$cm$^{-3}$. This phenomenon, caused by both a depletion effect and a dopant deactivation is well described in several studies[19, 20] (see Supporting Information).

SiNWs with various geometries have been tested, but in the following, only results obtained with two different devices are presented (see Supporting Information for results with other devices). The first one termed hereafter N67MHz is 2µm-long, 35nm-wide. The gap between electrodes and the nanowire is 152nm. The second suspended SiNW termed hereafter N120MHz is 1.5µm-long, 35nm-wide. The gap between electrodes and the nanowire is 82nm. The thickness is 30nm for both SiNWs. Frequency response of other devices are presented in Supporting Information.

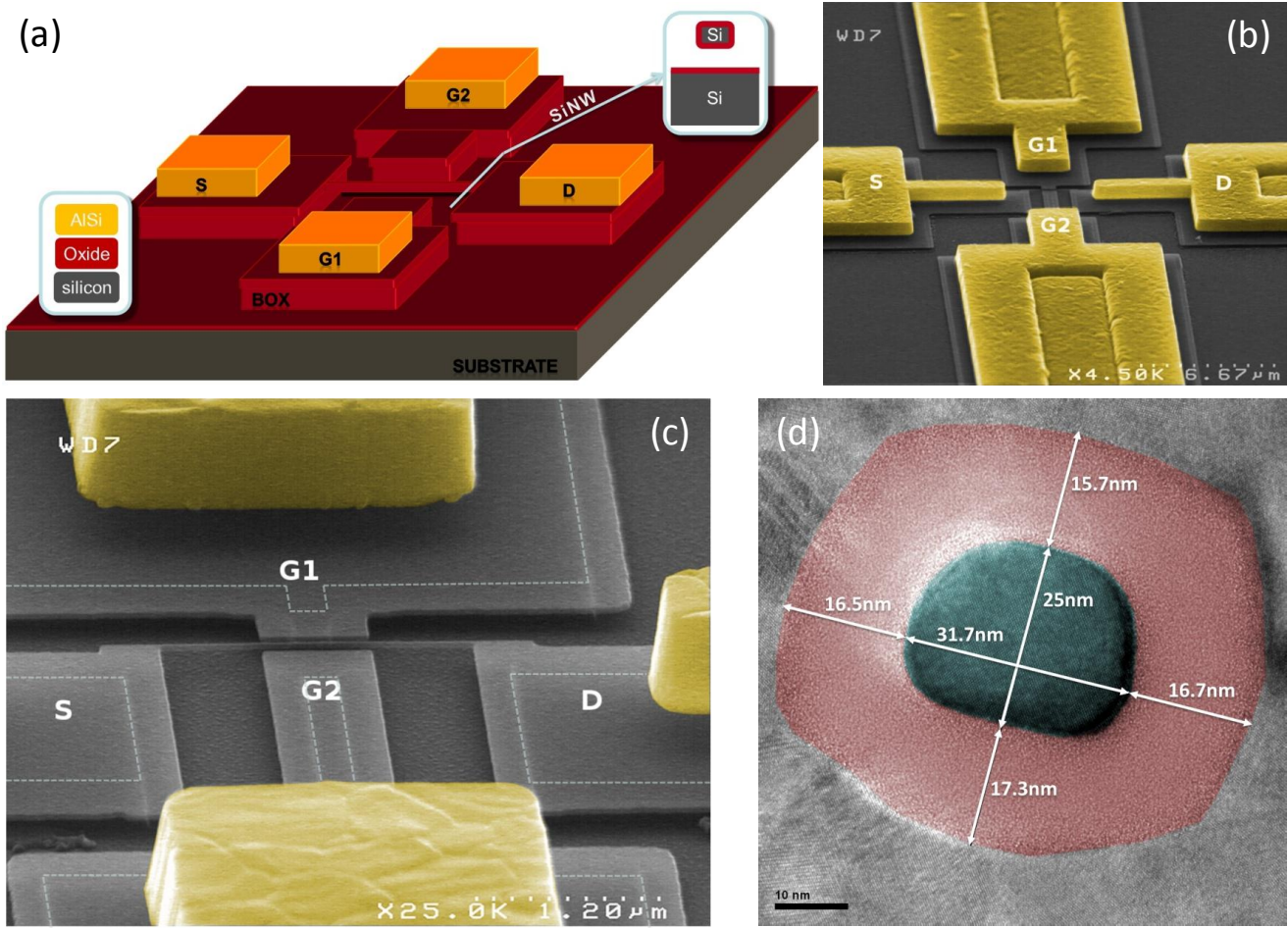

**Figure 1: Suspended junction-less Si nanowires with gates electrodes, in lateral field effect transistor (FET) configuration.** (a) 3D-schematic of the suspended SiNW showing layers: red for oxide, grey for silicon and yellow for metal. Lengths vary from 800nm to 3μm, widths from to 80 nm to 20 nm, gaps from 250nm to 80nm. Typical cross sections ranged from 20×20 nm² to 30×40 nm² depending on the use of oxidation/deoxidation steps for rounding the nanowire shape. (b) Scanning electron microscope (SEM) image of a SiNW resonator with two actuation electrodes (G1 & G2). (c) Zoom in on the SiNW – the white dotted line shows the undercut of the released buried oxide. (d) Transmission electron microscope (TEM) cross-section of a typical nanowire (prepared by Focused ion beam (FIB) – the silicon cross-section has dimensions of 32nm × 25nm.

The mechanical motion of such SiNWs modulates their electrical resistance through the piezoresistive effect. The extraction of SiNW PZR gauge factors has been discussed elsewhere, for example in [20]. Gauge factors $G$ between 80 and 240 – depending on the doping concentrations and surface states - have been extracted for similar SiNWs.. Moreover, static field-effect measurements will be presented as evidence of conductance modulation by electrostatic control[21]. This effect has been first presented for a suspended vibrating body by Bartsch *et al.* [16]. The possibility of controlling the current inside the nanowire with two uncorrelated approaches allowed us to perform resonance detection with two different self-transducing techniques.

## Piezoresistive (PZR) Effect

The PZR effect consists in the modulation of the resistance by application of mechanical strain on the device – that is the case for the PZR NW[14]. In fact, mechanical strain changes both the cross-section of the nanowire and the band curvatures resulting in a different conductivity inside the SiNW[20]. For the semiconductor nanowires, the contribution of the resistivity variation dominates over that due to the geometric change, and the relative resistance variation can be written as,

$$\frac{\Delta R}{R_0} = \frac{\Delta \rho}{\rho_0} = \varepsilon_l \cdot G \qquad (1)$$

where $\varepsilon_l$ is the longitudinal strain and $G$ the gauge factor.

Gauge factors extracted from our four-point bending tests are close to 240 as observed previously[20] for the two main devices tested in the work (N67MHz & N120MHz) – the measurement method is detailed in Ref [20].

The second order PZR effect for resonating doubly-clamped nanowires can be used for efficient transduction: the axial strain is then proportional to the square of the nanowire displacement (the first order stress field due to flexion is symmetric in a cross section, and hence the overall first order resistance is cancelled out)[14]

$$\frac{\Delta R}{R} = G \frac{\pi^2}{4} \left(\frac{a}{L}\right)^2 \qquad (2)$$

$L$ is the nanowire length. $a$ is the nanowire amplitude at the centre: $a(\omega) = z_1(\omega)\phi(\frac{L}{2})$ ($z_1(\omega)$ is the Fourier transform of the temporal part and $\phi(L/2)$ is the modal displacement).

The output current $I_{ds}$ is measured when applying a drain-source voltage $V_{DS}(\omega)$ and can be expressed as follows:

$$I_{DS}(\omega) = \frac{V_{DS}(\omega)}{R_{DS}(\omega)} = \frac{V_{DS}(\omega)}{R_0}(1 - \frac{\Delta R}{R_0}) = I_{DS0}(\omega) + I_{ds}(\omega)$$

$$I_{ds}(\omega) = \frac{V_{DS}(\omega)}{R_0} G \frac{\pi^2}{4} \left(\frac{a(\omega)}{L}\right)^2$$

(3)

The dynamic output current $I_{ds}(\omega)$ can be expressed with respect to the drain voltage $V_{ds}=V_{Bias}\cos((\omega-\Delta\omega)t)$ and the actuation signal applied on the gate $V_{Actuation}= V_G+ V_{AC}\cos(\omega t)$ in a down-mixing scheme (the SiNW acts like a non-linear mixer between actuation and bias). The NW's transverse displacement is proportional to the square of the actuation voltage and the drain current scales like $V_{Bias}$ and $V_{AC}^2$. Close to the resonance frequency the output current is

$$I_{ds}(\omega_0) = G \frac{V_{Bias}}{R_0} \frac{\pi^2}{4L^2} \frac{Q^2}{m^2\omega_0^4} \frac{(C_n\varepsilon_0)^2 L^2 t^2 V_G^2 V_{AC}^2}{g^4}$$

$$I_{ds}(\omega_0) \propto V_{Bias} \times V_{AC}^2$$

(4)

where $g$ and $t$ are the electrostatic gap and the nanowire thickness respectively. $C_n$, $\varepsilon_0$ are the fringe effect corrective factor (see their computation in Supplementary Information), and the vacuum permittivity respectively. $Q$ is the quality factor, $m$ the effective nanowire mass ($m=0.75\rho Lwt$) and $R_0$ the nanowire resistance at zero strain.

From equation (4) it is noticed that the output is proportional to the bias; while it varies quadratically with the AC-actuation $V_{AC}$: $f$ being the actuation (excitation) frequency, the drain current $I_{ds}$ varies at $2f$. It should be noted here that the gauge factor can also be roughly estimated with the equation (4), by measuring the current at resonance if the factor $C_n$ is carefully computed (see PZR measurements below).

**Field-Effect Transduction**

The junction-less field-effect transistor (FET) configuration allows the modulation of the SiNW resistance by electrostatic control. The SiNW constitutes a simple resistance controlled by an actuation electrode doped at the same level as the nanowire. The SiNWs are surrounded by a $SiO_2$ layer of 10nm, with an air gap of some tens of nanometers between the nanowire and the actuation electrode. The application of a DC voltage between the gate – and actuation electrode – modulates the number of free carriers in the nanowire resulting in a variation of the resistance. For p-type nanowires, increasing the gate voltage (for $V_G>0$) will gradually deplete the device thus reducing the current. The width $d$ of the depleted region can be calculated from the following equation[22]

$$d = \frac{-\varepsilon_s}{C_i} + \sqrt{\left(\frac{\varepsilon_s}{C_i}\right)^2 + \frac{2\varepsilon_s}{eN_a}(V_G - V_{FB} - V_{DS})} \quad (5)$$

where $\varepsilon_s$ is the permittivity of silicon, $N_a$ is the dopant concentration, $e$ is the electron charge ($1.6\times10^{-19}$ C), $C_i$ is the insulator capacitance including the oxide surrounding the nanowire and the air gap between the wire and the gate electrode. $V_G$, $V_{FB}$ and $V_{DS}$ are the gate voltage, the flat-band voltage and the drain voltage respectively (equal to $V_{Bias}$ in DC mode). The flat-band voltage depends on both the difference in work function between the gate and channel and the charge trap density at the silicon/oxide interface and in the oxide, $Q_{ss}$[22].

The resistance per unity of length of the nanowire is then,

$$dR = \frac{\rho(T,N_a)dx}{t(w-2d)} \quad (6)$$

Under the assumption of a gradual channel, the differential source/drain current can be deduced from the resistance from,

$$dI_{DS} = \frac{dV_{DS}}{dR} \quad (7)$$

The bulk current through the nanowire is obtained by integration of equation (7) along the channel. After a straightforward mathematical development an analytical expression of drain current, transconductance – in the linear regime – can also be found:

$$I_{DS}(V_G,V_{DS}) = \frac{2t}{\rho L}\left[\left(\frac{w}{2} - \frac{C_i}{eN_a}(V_G - V_{FB})\right)V_{DS} + \frac{C_i}{eN_a}\frac{V_{DS}^2}{2}\right]$$

$$g_m = \frac{\partial I_{DS}}{\partial V_G} = \frac{-2t}{\rho L}\frac{C_i}{eN_a}V_{DS} \quad (8)$$

Field-effect measurements have been performed in the static regime with a manual probe on the suspended nanowires N67MHz (with a dopant concentration of $3.8\times10^{17}$ cm$^{-3}$). Typical experimental characteristics $I_{DS}(V_G)$ for the N67MHz and N120MHz are plotted in Figure 2a) and b) respectively. The depletion mechanism for the device N120MHz is illustrated in Supplementary Information with COMSOL simulation based on the finite element method.

Theoretical drain currents obtained from equation (8) are also plotted in Figure 2 for a crude validation of the analytical model. The only parameters necessary for the drain current calculation are the surface charge density and the doping level. For the two suspended SiNWs we consider a charge density $Q_{ss}$ close to $2.5 \times 10^{11}$ cm$^{-3}$ which corresponds to the value found for similar fabrication process[20]. For the device N67MHz the doping level is set at $4 \times 10^{17}$ cm$^{-3}$ which is the experimental value. For the N120MHz the dopant concentration is set at $6 \times 10^{17}$ cm$^{-3}$, slightly larger than the experimental value for a good fit with measurements. The air gap capacitance $C_i$ and the hole mobility $\mu_p$ have to be known for a reasonable fit of the experimental data. $C_i$ computation is based on the simple parallel plate capacitance model but includes fringe effect: the parallel plate capacitance value is multiplied by a magnification factor $C_n$ that is evaluated at 1.95 and 1.5 for the devices N67MHZ and N120MHz, respectively. $\mu_p$ is computed around 200 cm$^2$V$^{-1}$s$^{-1}$ close to the mobility in bulk silicon (see Supplementary Information). Equation (8) gives fairly good trends and right orders of magnitude of the drain current for all data sets. In many cases the model fits well the experimental curves for low $V_G$ since it overestimates the experimental drain current for larger $V_G$. It may be due to the reduction of hole mobility at large electric fields. Additional current variation may be caused by the PZR mechanism since the SiNW undergoes stress when in motion. However the static displacement does not overcome 1nm for the voltages used here. The induced axial stress remains close to zero averaged over its cross section and hence the PZR contribution to the drain current remains negligible.

The theoretical and experimental transconductance $g_m$ of the device N120MHz are close to $3 \times 10^{-7}$S and $10^{-7}$S for $V_{DS}=-4.5$V and 1.5V respectively (see Figure 2c)). These values are similar to the values obtained with the suspended Fin-FET transistors[15].

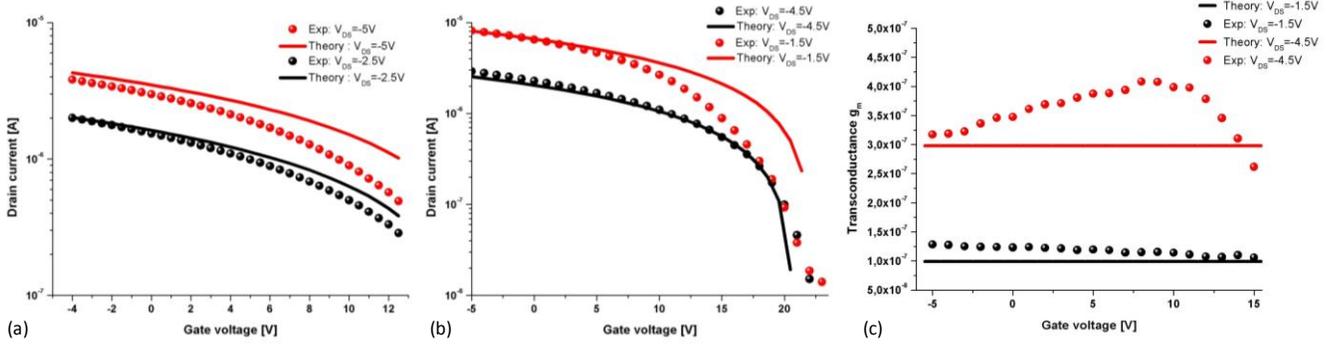

**Figure 2: SiNW field-effect transistor characteristics.** (a) $I_{DS}(V_G)$ characteristics for different drain voltages for the nanowire N67MHz – log plot. Increasing the gate voltage leads to depletion of the conduction channel and hence to a decrease of the passing current. (b) $I_{DS}(V_G)$ characteristics for different drain voltage $V_{DS}$ for the suspended SiNW N120MHz – log plot. (c) Transconductance $g_m$ according to $V_G$ for the device N120MHz: Here the gate voltage is set to reach a stronger depletion. The current is divided by a thousand when increasing the gate voltage from 0V to 25V. Above 25V, a leakage current of 1nA seems to appear. The nanowires used here have a similar structure to MOSFETs with the actuation electrode corresponding to the gate, the oxide corresponding to the dielectric and the nanowire to the transistor channel.

When performing resonance measurements, the suspended SiNW is actuated by a dc and an ac signal applied to the gate. The dc voltage will deplete the NW in a constant way, while the ac signal will deplete the NW in a dynamic mode. The effective gate voltage depends on the capacitance bridge formed by the air capacitance that is modulated by the beam displacement and the oxide capacitance around the NW. Knowing the transconductance $g_m$ the current variation caused by nanowire oscillations can be expressed according to actuation and bias voltages to obtain a similar expression as in (4). Keeping the same notation for actuation and bias voltages, close to the resonance frequency the output current can be expressed as,

$$I_{ds}(\omega_0) = g_m \delta V_G(\omega_0) \approx g_m V_G \frac{\delta C(\omega_0)}{C_{gap}} = g_m V_G \frac{a(\omega_0)}{g}$$
$$I_{ds}(\omega_0) = g_m V_G \frac{4}{3} \frac{(C_n \varepsilon_0 Lt) V_G V_{AC}}{g^3} \frac{Q}{m\omega_0^2} \quad (9)$$
$$I_{ds}(\omega) \propto V_{Bias} \times V_{AC}$$

where $C_{gap}$ is the air gap capacitance. The oxide capacitance is neglected as it is small and in series. Other parameters have been defined elsewhere in the text.

Equation (9) shows how the nanowire resistance is modulated by the field effect, which is a first order phenomenon as opposed to the PZR transduction described above. With $f$ being the bias frequency, the nanowire has therefore to be biased at near 1×$f$ for the field-effect transduction while it has to be biased at 2×$f$ for the PZR transduction. A simple set-up can hence be used with little change for each transduction. It should be noted that the capacitive dynamic current (roughly estimated from $\varepsilon_0 S/g^2 \omega_0 a(\omega_0)$) is between 0.1pA and 0.01pA. It is two to three orders of magnitude lower than the experimental current presented in the next section).

## High Frequency Experimental Results

### Experimental Schemes and Techniques

High frequency electromechanical measurements have been performed in a probe station at room temperature for both transduction schemes on the same SiNWs. The readout schemes are based on heterodyne techniques (see **Erreur ! Source du renvoi introuvable.**a) & Figure 4a)) previously used for SiNW PZR detection[14] or field-effect transduction of carbon nanotubes[23].

In the case of the FET transduction, the transconductance $g_m$ is modulated by the bias voltage $V_{Bias}$ at $\omega-\Delta\omega$ since the displacement varies with the AC-actuation $V_{AC}$ at $\omega$. The nanowire acts as a high frequency (HF) mixer – see equation (10) – and the drain current varies at $\Delta\omega$. In the case of the PZR transduction the down-mixing scheme is similar but with an AC-actuation (excitation) at $2\omega$ – see equation (4). For the sake of clarity we will refer to 1×$f$ when the natural frequency of the nanowire is used and 2×$f$ for twice its natural frequency.

The biasing and actuation voltages used to operate in the linear regime are similar for the two transduction schemes. $V_{Bias}$ is set between 10mV and 150mV. The AC-actuation $V_{AC}$ is set between 50mV and 200 mV since the DC voltage $V_{DC}$ is around a few hundreds of millivolts – 300mV in the example below (*i.e.*, $V_G$=300mV for the FET detection). The use of similar voltage values allows a direct comparison of the signal-to-noise (SNR) and the signal-to-background (SBR) ratios obtained with both techniques. The noise spectral power density is also measured for the two transduction schemes to further characterize the noise sources in each case.

## Resonance Measurements

Results obtained with the PZR transduction show a quadratic variation of the peak amplitude with the actuation voltage (**Erreur ! Source du renvoi introuvable.**b), while variation with the bias remains linear (**Erreur ! Source du renvoi introuvable.**c) as expected from equation (4).

As explained above the gauge factor can be estimated from the dynamic drain current by using the equation (4). For the operating point ($V_{DC}$=0.3V, $V_{AC}$=0.1V, $V_{Bias}$=0.07V), the drain current at resonance is 10.25 pA. Knowing the initial resistance ($R_0$ = 1.2 MΩ), the actual sizes (SiNW section=35nmx30nm and length=2μm) and the resonance frequency (67.25 MHz) with a *Q*-factor of 700, the displacement is evaluated at about 1nm (peak) for an axial stress around 150 kPa. The gauge factor is then close to 200 which matches well with extracted values in static regime[20].

The theoretical current amplitudes are calculated with the electrical parameter values determined in the previous section (*i.e.*, $C_n$=1.95, $Q_{ss}$=1.5x10$^{11}$ cm$^{-2}$, $N_a$=4x10$^{17}$ cm$^{-3}$) and with a gauge factor of 200. The model shows the right order of magnitude as well as correct trends.

On the contrary results obtained with FET-transduction show a linear variation of the resonance amplitude when increasing actuation and bias voltages (see Figure 4b) & Figure 4c) respectively) as predicted by equation (9). The theoretical current amplitude is calculated assuming the same electrical parameters as for the PZR transduction. The *Q*-factor extracted from the measurement is set to 700 and the resonance frequency is set at 67.27MHz. $g_m$ is calculated with Equation (8). The theoretical results are in a good agreement with the experimental values, showing that a simple model can predict the order of magnitude of the electromechanical response.

These results have been obtained on the same device. The SBR for the FET transduction is of the order of 2.5 depending on polarization conditions whereas the SBR in the case of the PZR transduction is of the order 4.5. Results obtained with various devices are shown in Supplementary Information.

Systematic measurements lead to the conclusion that the PZR detection seems to be slightly (about 50%) more efficient than the FET-detection scheme.

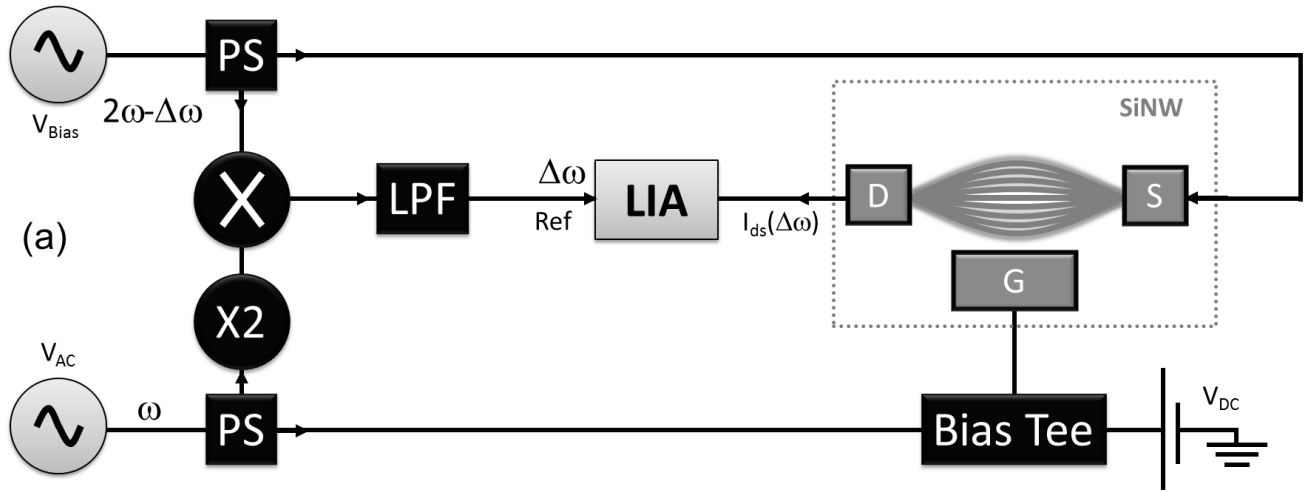

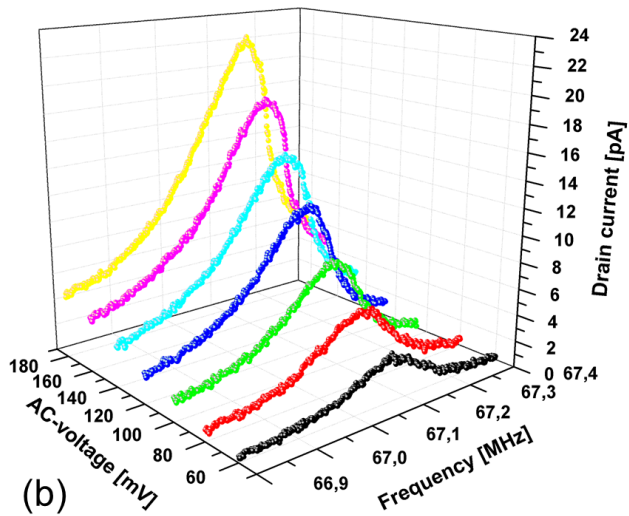

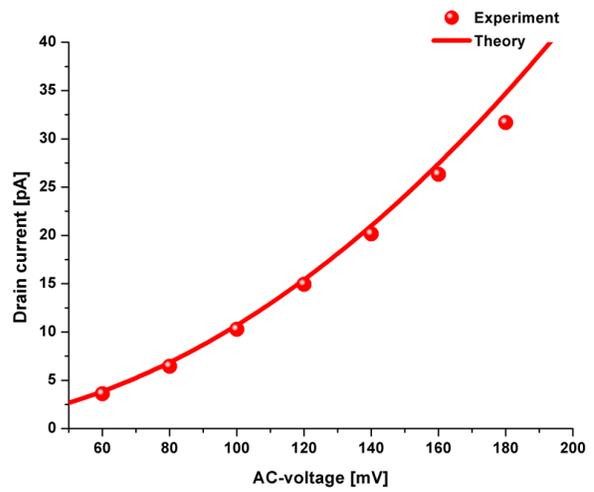

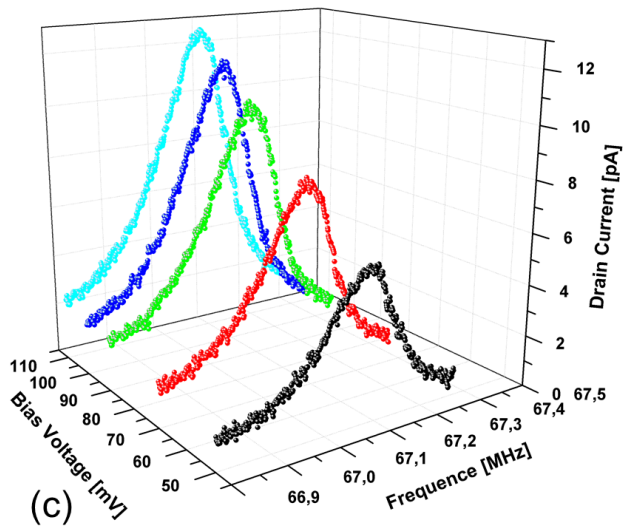

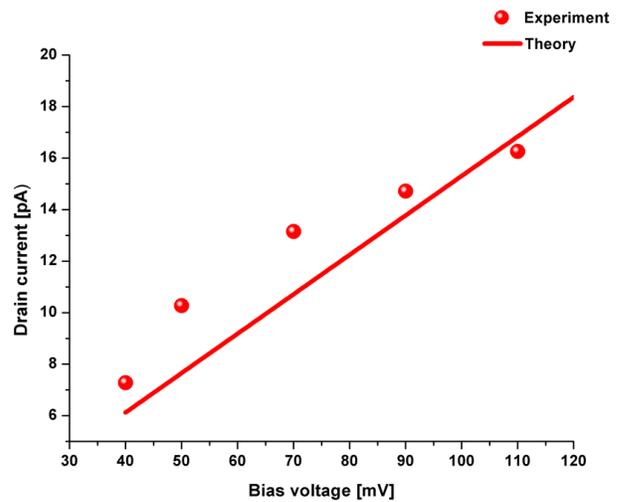

**Figure 3: Piezoresistive transduction of SiNW resonators**. (a) Readout scheme used for PZR transduction. The nanowire is biased with a $2\omega$ signal as the resistance variation due to the second-order PZR effect, while actuation is performed at $1\omega$. The output signal is detected with a lock-in amplifier (LIA). A low pass filter (LPF) is used to reject parasitic frequencies coming from mixer and RF synthesizers. RF-signals are split with two power splitters (PS) – (b) Electromechanical resonances for various actuation voltages $V_{AC}$. The amplitude varies quadratically with the actuation voltage as expected from theory. – (c) Electromechanical resonances for various bias voltages $V_{Bias}$. The amplitude is linear with the bias voltage.

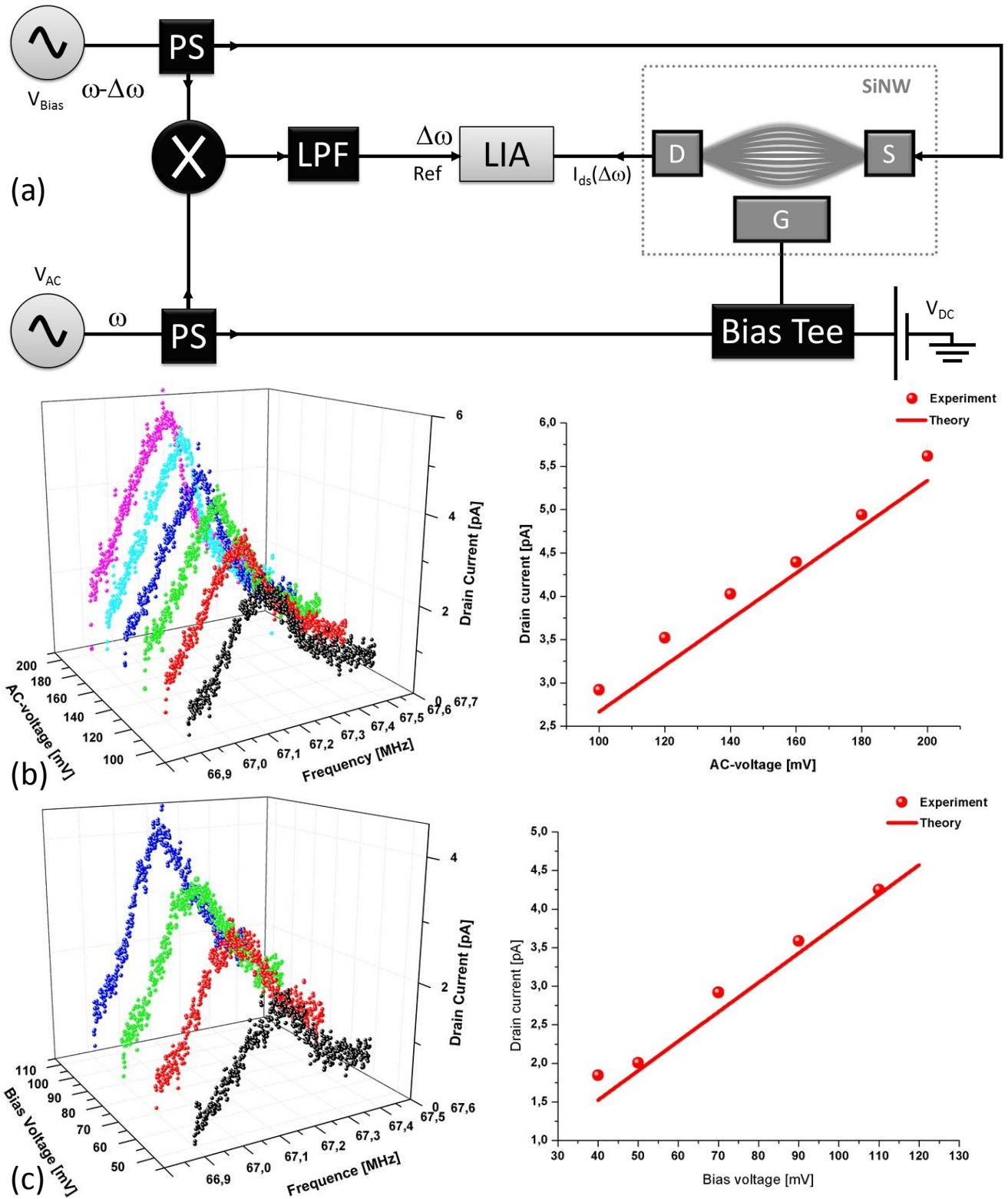

**Figure 4: High frequency transduction of SiNW resonators via the embedded field effect transistor.** (a) Readout scheme used for field-effect transduction of the resonant nanowire. In this case, both actuation and bias are 1ω-signals. The pick-up of the output signal is performed with a LIA as in the PZR case – (b) Electromechanical resonances for different actuation voltages $V_{AC}$. The amplitude variation is proportional to the actuation voltage as expected from theory. – (c) Electromechanical resonances for various bias $V_{Bias}$. The amplitude is linear with the bias voltage.

As discussed in previous sections, one of the main motivations here is the development of a smart, compact SiNW pixel unit for potential resonant mass sensing with such pixel arrays. Like for any other application involving resonant detection, the frequency stability $\delta f/f_0$ of such a pixel is a key property. One usual estimation of this frequency stability is the Allan deviation[24] $\sigma_A$ versus an integration time $\tau$. For dominant amplitude white noise and narrow integration bandwidth $1/\tau$ compared to the eigen-frequency $\omega_0$, $\sigma_A$ follows an asymptotic law:

$$\frac{\delta f}{f_0}(\tau) = \frac{1}{2Q}\frac{\sqrt{1/\tau}}{SNR} \tag{11}$$

The frequency fluctuation measurements have been performed on the N67MHz resonator with both transduction mechanisms. The Allan deviation is measured in open loop recording the phase variation of the drain current[10].

Allan deviation results of the two transduction schemes are presented Figure 5, with integration times in a range of practical use for most applications. They are following a $\tau^{-1/2}$ power law, in agreement with equation (11). This shows that white noise processes are dominant over this whole range of integration times. In the FET-detection case, the slope seems to inflect at 1s-integration time due to $1/f$-noise appearance. In the studied case, the thermomechanical noise is estimated to 0.044 pm/√Hz, which corresponds to 1.13 fA/√Hz and 22 fA/√Hz for the PZR and FET transduction respectively, thus remaining negligible. The Johnson noise is the second obvious white noise process. Its magnitude is estimated to 0.043 pA/√Hz by measuring its spectral power density with a PXI (Peripheral Component Interconnect for Instruments) card. The lock-in amplifier exhibits a white noise density of 0.13pA/√Hz, which is the dominant noise process in the measurement scheme. As the PZR-detection is more efficient – current amplitude of 17.5 pA against 4.5 pA for the FET-detection for the same operating points – it offers an improved short-term frequency stability versus which obtained with a FET-transduction (see equation (11)). The best Allan deviation in our time range is close to 3ppm with the PZR-transduction while it reaches up to 20ppm with the FET-transduction. This is also the case for the nanowire resonating at 145.5MHz shown in SI. The Allan deviation results obtained with the PZR-transduction on many devices are comparable to results presented by Bachtold *et al.*[8], with frequencies in the GHz range, but at liquid helium temperatures.

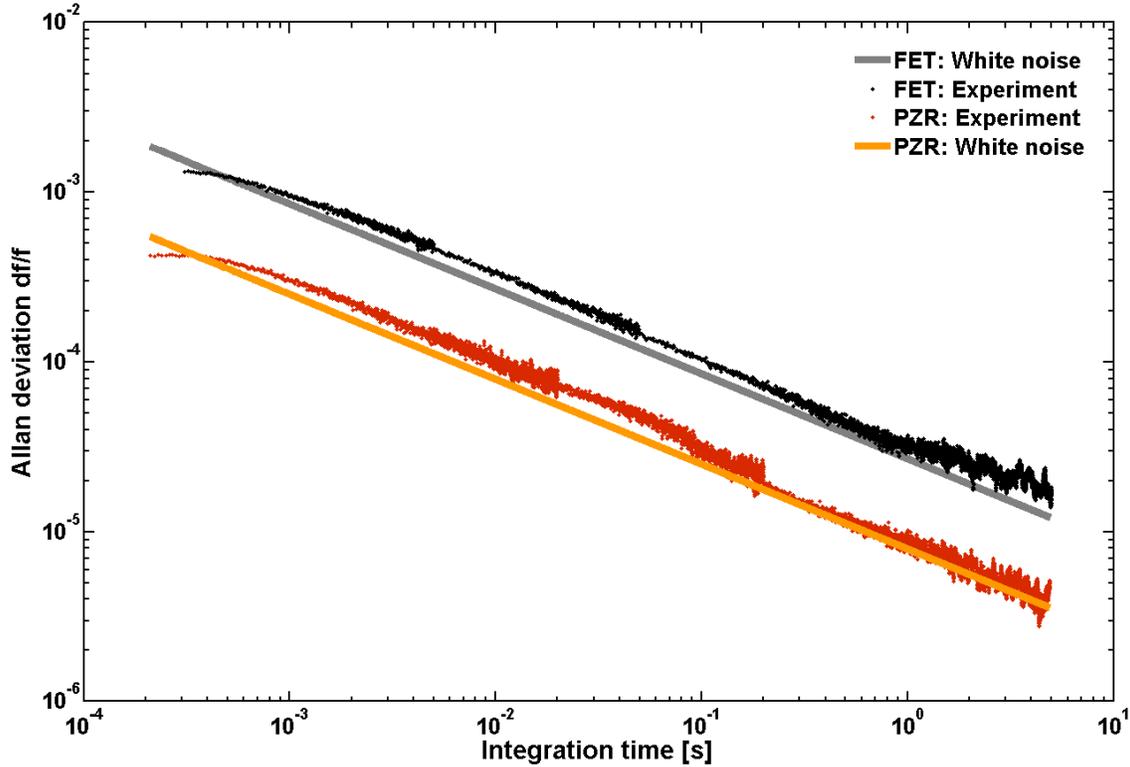

**Figure 5: Frequency stability (Allan deviation) measured in both PZR and field-effect transduction schemes**. The suspended SiNW is driven at its resonant frequency at the onset of the non-linearity (resonance signals are 17 pA and 4.5 pA for PZR and FET transductions respectively). The Allan deviation is measured in three steps (for short, intermediate and long times). For short times (<0.05), the integration time of the lock-in and the global acquisition time are 100 μs and 1s respectively. For intermediate times (0.05s <$\tau$< 0.5s), they are set to 100μs and 10s respectively. For long times (0.5 <$\tau$<5s) they are set to 1ms and 100s respectively. These adjustments remove the effect of the lock-in filtering that would artificially decrease the Allan deviation and ensure at least 20 points for each interval. They follow a clear $\tau^{-1/2}$ power law, which shows that white noise processes are dominant over the whole range of integration times. The best Allan deviation in this range is close to 3 ppm with the PZR-transduction while it reaches up to 20ppm with the FET-transduction.

In conclusion, we have experimentally demonstrated some of the smallest top-down SiNW NEMS resonators, with minimum width down to ~20nm, and resonance frequencies up to 150MHz, in the mid-upper very high frequency (VHF) band. This was achieved thanks to a top-down, 200mm VLSI process fully amenable to monolithic CMOS co-integration, as was demonstrated elsewhere with first-order PZR devices[18,25]. The very small dimensions have allowed for the first realization of a vibrating SiNW NEMS, where the conductance is controlled by a nearby gate without junction. The electrical performance of such field effect detection has been compared to the self-transducing piezoresistive scheme within the same SiNW. Both schemes have shown similar SBRs, and the PZR transduction has exhibited a slightly higher output signal. As the dominant source of noise in both cases is the input noise of the LIA, independent from the transduction type, the PZR one shows a correspondingly better SNR. Frequency stability of the same SiNW has been measured with both transductions, and we obtain 20ppm for the FET detection and 3ppm for the PZR detection at ambient temperature and low pressure. Piezoresistive gauge factors have been measured with two distinct methods (static and dynamic regimes), which agree well.

In the near future, the junction-less field effect transconductance will be improved thanks to efforts on the process and geometry (smaller widths and gaps, lead resistance…), as well as on the doping level, at the expense of the gauge factor. Both transduction schemes should hence show very close SNRs; one could consider using both transductions *simultaneously* to increase the output signal, by using two bias harmonic at the same time. CMOS co-integration, already demonstrated with this very same process, will undeniably boost the SNR of both transductions. Based on our experience, we are confident that Allan deviation of $\sigma_A$~1ppm can be reasonably reached with such a device (which corresponds to the level set by the Johnson noise). Considering the device effective mass, this translates into a sub-zg (1zg=$10^{-21}$g) mass resolution at ambient temperature and a few seconds integration time. An order of magnitude improvement is to be expected when operating at low temperatures, yielding a limit of detection of the order of 50yg (1yg=$10^{-24}$g). Only bottom-up devices like graphene nano-ribbons[26] or carbon nanotubes[8] have reached similar or better mass resolutions. These performances combined with a VLSI, CMOS compatible process open up new sensing paradigms: the compact, simple topology of this elementary pixel is ideally suited for ultra-dense arrays. This is of paramount importance for increased capture cross-section for gas sensing[13] or real-world NEMS mass spectrometry systems[12]. Also, a variety of applications within the More-Than-Moore framework will benefit from the array topology, like logic switches[5] or RF receivers[27] where each NEMS is a channel; if noise processes are uncorrelated, an N-device array will provide a $\sqrt{N}$ increase in SNR, which is particularly interesting for time keeping [4] for instance.

ACKNOWLEDGMENTS

The authors acknowledge financial supports from both the European Commission under the FP7 STREP NEMS-IC program under the Grant Agreement N°224525. P.X.-L.Feng thanks partial support from Case School of Engineering. We thank Carine Marcoux for her helpful work on the fabrication technology.